\documentclass[aps,prl,twocolumn,showpacs,superscriptaddress,floatfix,amsmath,amssymb,longbibliography]{revtex4-2}

\usepackage[titletoc,toc,title]{appendix}
\usepackage{graphicx}
\usepackage{bm}
\usepackage{xcolor}
\usepackage{booktabs}
\usepackage{hyperref}
\usepackage{diagbox}
\usepackage{braket}

\renewcommand{\vec}[1]{\bm{#1}}

\begin{document}

\title{Signatures of electronic correlations and spin-susceptibility anisotropy in nuclear magnetic resonance}

\author{Stephen Carr}
\affiliation{Department of Physics, Brown University, Providence, Rhode Island 02912-1843, USA}
\affiliation{Brown Theoretical Physics Center, Brown University, Providence, Rhode Island 02912-1843, USA.}
\author{Charles Snider}
\affiliation{Department of Physics, Brown University, Providence, Rhode Island 02912-1843, USA}
\author{D. E. Feldman}
\affiliation{Department of Physics, Brown University, Providence, Rhode Island 02912-1843, USA}
\affiliation{Brown Theoretical Physics Center, Brown University, Providence, Rhode Island 02912-1843, USA.}
\author{Chandrasekhar Ramanathan}
\affiliation{Department of Physics and Astronomy, Dartmouth College, Hanover, NH 03755, USA}
\author{J. B. Marston}
\affiliation{Department of Physics, Brown University, Providence, Rhode Island 02912-1843, USA}
\affiliation{Brown Theoretical Physics Center, Brown University, Providence, Rhode Island 02912-1843, USA.}
\author{V. F. Mitrovi\ifmmode \acute{c}\else \'{c}\fi{}}
\affiliation{Department of Physics, Brown University, Providence, Rhode Island 02912-1843, USA}

\date{\today}

\begin{abstract}
We present a methodology for probing the details of electronic susceptibility through minimally-invasive nuclear magnetic resonance techniques.
Specifically,  we classify electron-mediated long-range interactions in an ensemble of nuclear spins by revealing their effect on simple spin echo experiments.
We find that pulse strength and applied field orientation dependence of these spin echo measurements resolves the spatial extent and anisotropy of electronic spin susceptibility.
This work provides an alternate explanation to NMR results in superconducting and magnetically-ordered systems.
The methodology has direct applications for sensing and characterizing emergent electronic phases.
\end{abstract}

\maketitle



Nuclear magnetic resonance (NMR) traditionally measures dissipation and/or loss of coherence  due to  spatial redistribution of the nuclear spins, with a focus on the temperature dependence of the spin-lattice relaxation  rates ($T_1$), 
spin-spin relaxation rates ($T_2$), or NMR shift ($K$) ~\cite{Pennington1989, Pennington1991, Horvatic1993, Mitrovic2002}.
Sudden changes in the dissipation rates can be compared to models for electron-nuclear or  phonon-nuclear interactions, allowing for the microscopic observation of electronic phases.
This standard approach for NMR as an experimental probe relies on spin decoherence and provides relationships between electronic spin-susceptibility at high symmetry points and the nuclear $T_1$ or $T_2$ rates.
NMR is an attractive tool for probing electronic ground state properties as it uses low frequency excitations relative to electronic energies.
Recently identified quantum phases of matter may encode details of their  intricate structure into NMR responses in ways that lay outside this current paradigm.

It is not uncommon in NMR studies of strongly correlated materials to observe unusual time-asymmetric features under standard spin echo protocols~\cite{VesnaNote}.
These are typically classified as experimental artifacts, often attributed to an uncontrolled phase transition as strong RF pulses can cause electronic heating in the sample~\cite{Ishida2019, Pustogow2019, Vinograd2021, Mitrovic2008}.
To provide an alternate explanation for these unconventional signals, we investigate the time evolution of nuclear spins with electron-mediated interactions on a 2D lattice.
When the interaction is anisotropic, clear signatures emerge during pulse angle sweeps.
The radial form and range of the interaction is also partially recoverable from careful analysis of the spin dynamics.
As the details of the nuclear interaction are inherited from the electronic spin susceptibility, one can determine many features of electronic spin-spin correlation previously inaccessible by NMR.
In this letter we demonstrate a unique paradigm to extract range and anisotropy of electronic spin correlations through a series of simple NMR experiments.

\begin{figure}[ht]
\centering
\includegraphics[width=\linewidth]{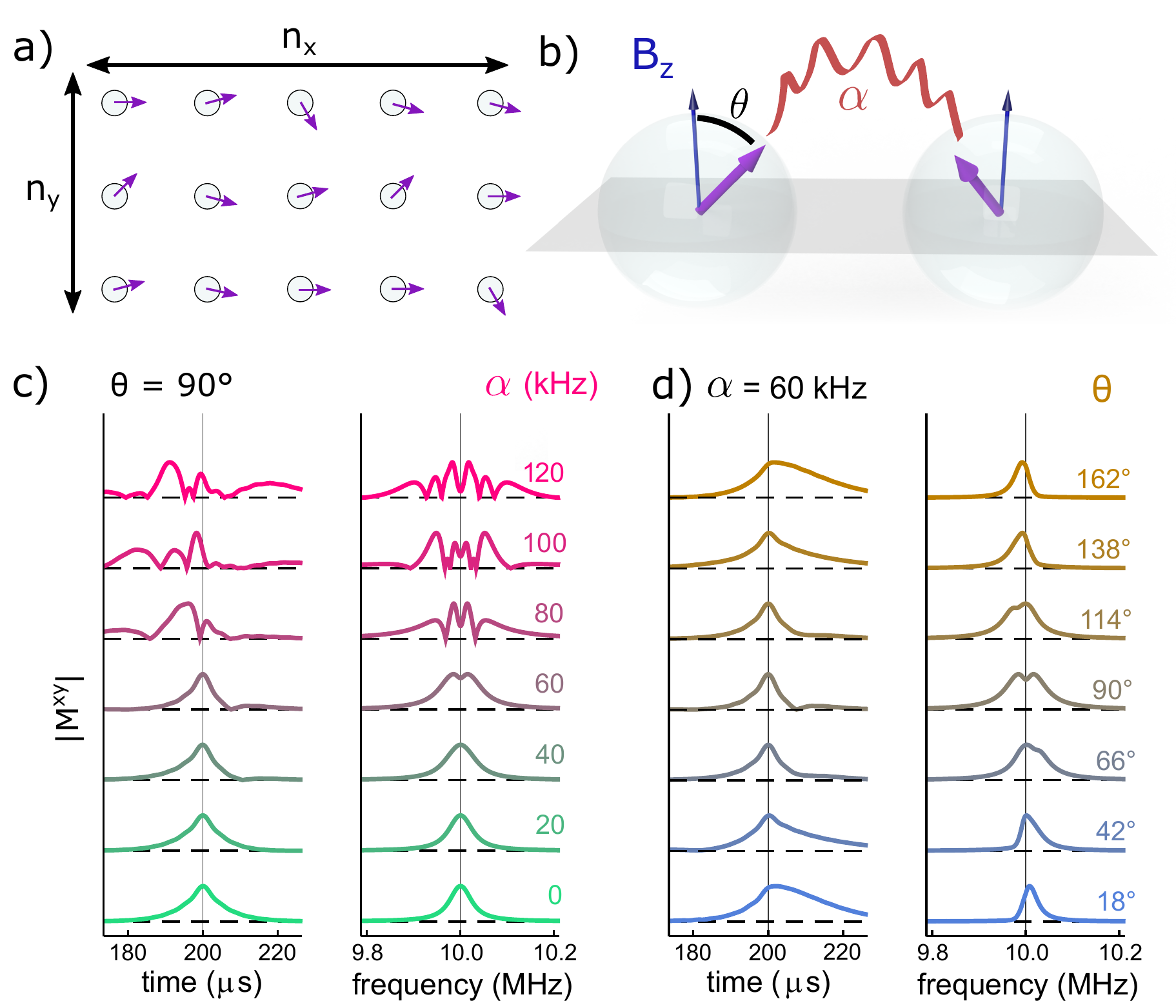}
\caption{\textbf{a)} Top-down view of a square lattice of nuclear spins of size $n_x \times n_y$.
\textbf{b)} The interaction between nuclei depends on a characteristic strength $\alpha$. By varying the strength of the first applied pulse the expected angle $\theta$ between the spins and the constant $B_z$ can be modified.
\textbf{c)} Average planar magnetization from spin echo simulations, in both the time and frequency domain, for a fixed $\theta$ and varying $\alpha$, and \textbf{d)} vice-versa.
}
\label{fig:power_sweep}
\end{figure}


Long-range interactions between magnetic particles through conduction electrons have been studied previously~\cite{Ouazi2006, Sobota2007, Bermudez2011, Pruser2014, Mohammadi2020}, but a theory for an interacting lattice of spinors with long-range couplings is undeveloped.
The general model for coupled nuclear spins is well understood~\cite{Robert1982}, with many packages available for treatment of the full Hilbert space (limited to $N \approx 20$ spins)~\cite{Bak2000, Tosner2014, Veshtort2006, Hogben2011}.
Truncated Louisville space representations can handle up to $N \approx 1000$ spins~\cite{Dumez2010, Karabanov2015, Karabanov2021}, but we find that even this is not large enough to capture the emergent properties from long-range electronic correlations.
Classical treatments with only nearest-neighbor coupling have found good agreement with quantum methods~\cite{Tang1992}, and the agreement generally improves as the number of interacting neighbors increases~\cite{Elsayed2015}.


We consider tens of thousands of spins and treat the interactions at the mean-field level~\cite{Snider2021, SpinEchoCode}.
Our simulations are performed on an $n_x \times n_y$ square lattice of unit length with spin-$\frac{1}{2}$ nuclei (Fig.~\ref{fig:power_sweep}a) with periodic boundary conditions.
We perform small $\Delta t$ updates on each spin in the ensemble, with $\Delta t$ chosen small enough to prevent any error from approximation of the Hamiltonian matrix exponential during time propagation (see SM~\cite{SMnote}).
The spinors have a central frequency $\nu = 10$ MHz with small variations in the environment of each nuclei causing deviations, which we model as a Lorentzian with linewidth $\Gamma = 25$ kHz.
To achieve a spin echo, at $t=0$ a $\theta$-strength $I^x$ pulse brings all the spins out of alignment with $B_z$ and towards the $xy$ plane. After a time $\tau$, a $2 \theta$-strength $I^y$ pulse rotates them about the $y$ axis. When $\theta = 90^\circ$ this achieves a perfect $180^\circ$ rotation of the spins, cancelling the accumulated phases from the frequency variations and forming a spin echo at $t = 2 \tau$.
We present the in-plane net magnetization of the spin ensemble, $\bar{M}^{xy}(t) = \bar{M}^x(t) + i \bar{M}^y(t)$, and its Fourier transform (see SM) in Fig.~\ref{fig:power_sweep}c,d.
We use the notation $\bar{M}$ for the global net magnetization, to distinguish from a local magnetization, $M$.

Applying second order perturbation theory to the hyperfine interactions ($\Delta$) between electrons and nuclei leads to an effective spin-spin interaction bilinear in the nuclear spins and quadratic in the hyperfine strength~\cite{Ruderman1954, Yosida1957}.
The effective Hamiltonian for the spin-spin interaction between nuclei takes the form
\begin{equation}
\label{eq:H_perturb}
    H_I(i,j) \propto \Delta^2 \vec{I}_i^\dagger \chi(R_i - R_j) \vec{I}_j
\end{equation}
where $\chi$ is the spin susceptibility of the electrons and $\vec{I} = (I^x, I^y, I^z)$ are the nuclear spin operators.
The form of Eq.~\ref{eq:H_perturb} avoids assumption of an isotropic Fermi liquid, and also makes explicit the proportional relationship between $\chi$ and the nuclei-nuclei coupling.
Making a mean-field approximation of the interaction in Eq.~\ref{eq:H_perturb}, we obtain
\begin{equation}
\label{eq:H_MF}
H_{mf}(i) = -\nu_i I_i^z - \sum_{d=x,y,z} \alpha_d I_i^d M_i^d
\end{equation}
with $\nu_i$ the resonant frequency of the non-interacting spin, $M_i^d$ the mean magnetization along the $d$'th axis seen by a spinor at lattice site $i$ from the other spins, and $\alpha_d$ the effective strength of the hyper-fine electron-mediated coupling along that spin axis, as illustrated in Fig.~\ref{fig:power_sweep}b.
We treat the spin operators as unitless and absorb all relevant physical constants into $\nu$ and $\alpha$, which take units of Hz.

We expect the introduction of a term bilinear in the nuclear spin operator $\vec{I}$ to break the even symmetry of $|\bar{M}_{xy}(t)|$ around the spin echo.
The time-evolution of the spins can be estimated by $dH / d\vec{I}$, which acts as an effective torque on each spinor.
For the non-interacting case, $dH / d\vec{I} = -\nu \hat{z}$, a constant, and so if the initial distribution of spins is frequency-symmetric the resulting echo will be time-symmetric.
$\vec{M}(t)$ also acts as an effective torque however, and allows for the breaking of time-symmetry in the spin echo.


We begin with the simplest isotropic infinite-range interaction form, $\vec{M}_i = \vec{\bar{M}} = \sum_j \braket{\vec{I}_j}/N$ and $\alpha_d \equiv \alpha$.
This uses the net magnetization of the entire ensemble ($N$ spins) as the local magnetization when determining $H_{mf}$, leaving $\alpha$ and the pulse angle $\theta$ as the only unfixed parameters.
The role of the coupling strength $\alpha$ is investigated first in Fig.~\ref{fig:power_sweep}c.
Weak $\alpha$ values ($< 40$ kHz) show a nearly perfect spin echo in both the time and frequency domain. As $\alpha$ grows, time asymmetric echos occur.
The interaction causes the most significant changes to the spin evolution near the echo and shortly after the initial pulse (free induction decay, or FID).
For $\alpha < 70$ kHz, the only noticeable effect on $M(t)$ occurs near the echo time, showing up as a small post-echo shoulder.
At larger $\alpha$ values, the interactions cause significant ringing even during the FID (see SM).
In Fig.~\ref{fig:power_sweep}d, the effect of different pulse strengths on the spin echo are compared.
There are many reductions in the magnetization near 10 MHz reminiscent of spectral hole burning, so the signatures of strong electron-mediated nuclear coupling could easily be miss-attributed to over-pumping the system \cite{Mitrovic2008}.


\begin{figure}[htbp]
\centering
\includegraphics[width=\linewidth]{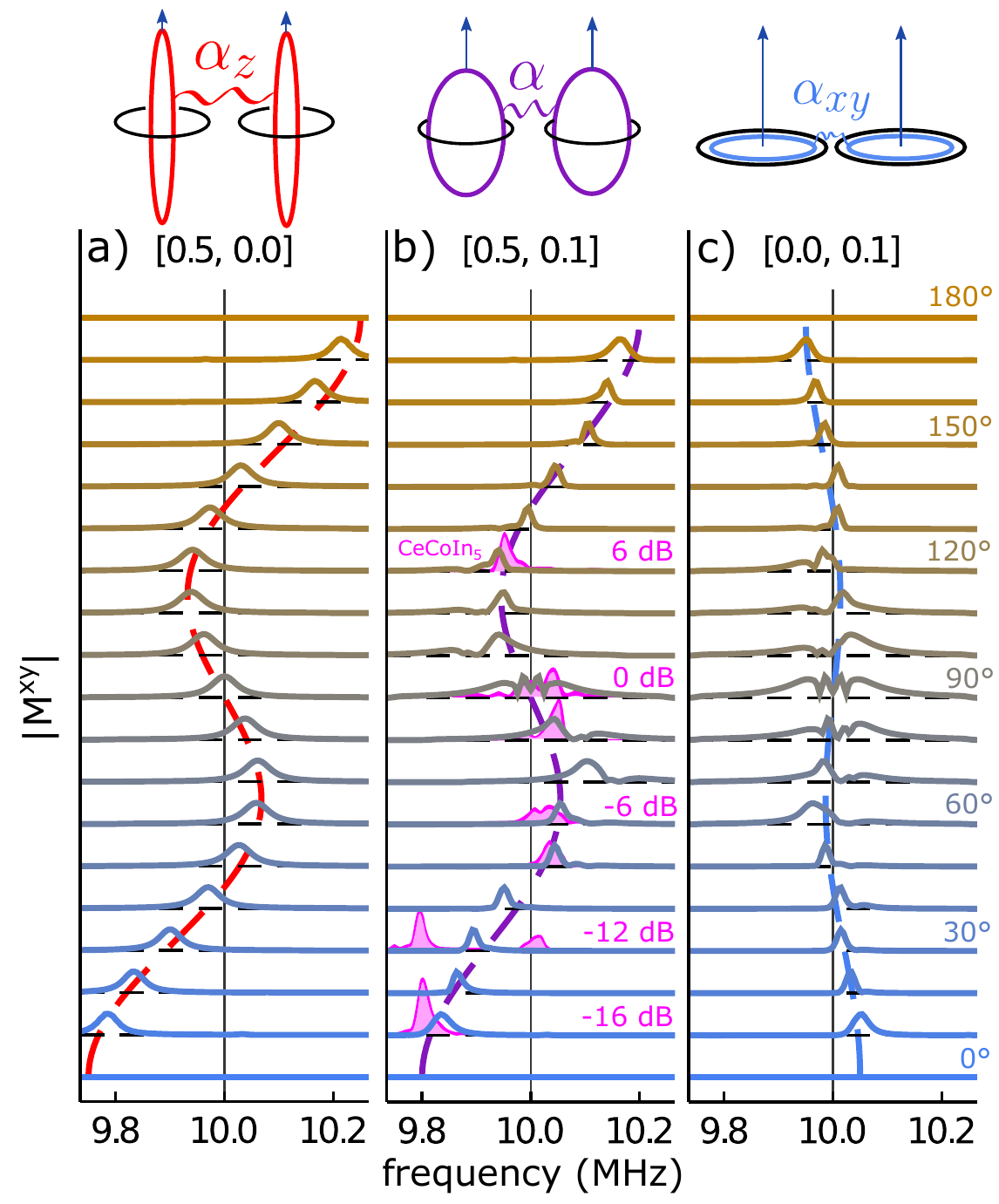}
\caption{Pulse angle ($\theta$) dependence of the NMR spectra for different aspect ratios of the effective spin-spin interaction,  [$\alpha_z$, $\alpha_{xy}$] in MHz. The S-function $( (\alpha_{xy} - \alpha_{z})/4) (\cos{\theta} + \cos{3 \theta})$ is given by the dashed line for each aspect ratio. In $\textbf{b)}$ the pulse-dependent spectra of $^{115}$In$(1)$ in CeCoIn$_5$ at $T = 70$ mK and $B = 9.55$ T is reproduced in pink from Ref. \onlinecite{Mitrovic2008} (pulse power given in dB).}
\label{fig:ratio_sweep}
\end{figure}

To remove the assumption of an isotropic interaction, we introduce the axis-dependent couplings, $\alpha_z \neq \alpha_{x} = \alpha_y \equiv \alpha_{xy}$,
motivated by anisotropy in the electronic spin susceptibility: $\chi^{zz} \neq \chi^{xx}, \chi^{yy}$.
This can occur in layered materials~\cite{Joy1992} or be caused by spontaneous electronic nematicity~\cite{Pomeranchuk1958, Kivelson1998, Halboth2000, Oganesyan2001, Fradkin2010}.
Fig.~\ref{fig:ratio_sweep} investigates three different conditions for the anistropic interaction: $\alpha_{xy} = 0$, $\alpha_{z} = 0$, and $\alpha_{xy} \neq \alpha_z$.
For $\alpha_{xy} = 0$ (Fig.~\ref{fig:ratio_sweep}a) the interaction simply introduces an additional $I^z$ term, increasing or decreasing the average resonant frequency of the ensemble.
Understanding the distribution of spins in the absence of interactions reveals how $\theta$ shifts the resonant frequency.

\begin{figure}[htbp]
\centering
\includegraphics[width=\linewidth]{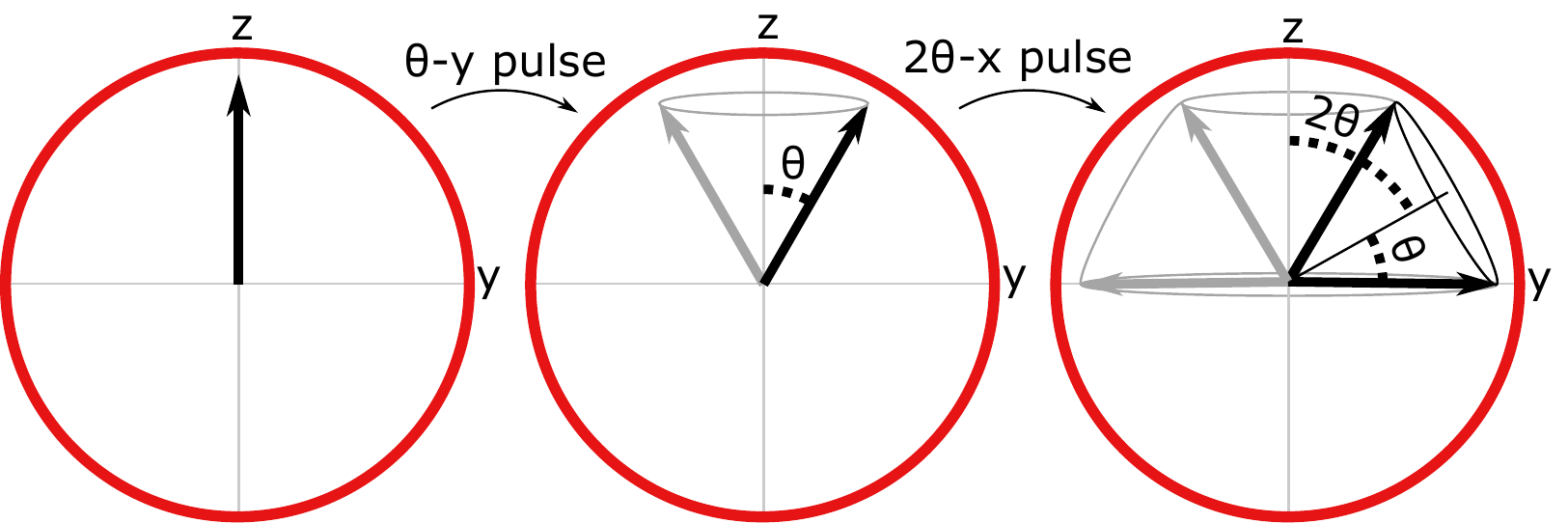}
\caption{Distribution of nuclear spins in the absence of the spin-spin interaction during a $[\theta$,$2\theta]$ pulse sequence. After each pulse, the updated spin distribution is given by a dark black line, and the distribution a short time after the pulse is given by the grey outlines.
The distribution just before the $2 \theta$ pulse traces out a cone, and so the last figure considers a rotated cone instead of a rotated arrow.
}
\label{fig:spin_diagram}
\end{figure}

We have derived the values of $\vec{\bar{M}}$ in the non-interacting case exactly in the SM, but here we outline the argument for $\bar{M}^z$ by representing the spins as a $SO(3)$ vector of magnitude $1/2$ (Fig.~\ref{fig:spin_diagram}).
The first $\theta$ pulse moves all the spins an angle $\theta$ off the $z$-axis, where they then precess because of $B_z$ and trace out a ring centered along the $z$-axis.
Assuming the time between each pulse ($\tau$) is long enough to ensure that the spins are uniformly distributed, the second $2 \theta$ pulse then rotates the now uniform ring of spins an additional angle $\theta$ away from the $z$-axis.
The average $z$-component of the spins just after the $2 \theta$ pulse is given by the average of the maximal and minimal $z$-component values of the tilted ring, $\cos \theta$ and $\cos 3\theta$ respectively. The $z$-component is unchanged under further time evolution by $B_z$.
Therefore $\bar{M}^z$ during the spin echo is $(\alpha_z/4) (\cos\theta + \cos 3\theta)$, which we denote as an S-function, and in agreement with the frequency shift observed in the simulations.
Although Fig.~\ref{fig:spin_diagram} only shows the case for $\theta < 90^\circ$, our derivation of the average $\bar{M}^z$ value holds for all $\theta$.

Considering instead $\alpha_z = 0$ (Fig.~\ref{fig:ratio_sweep}c), one can estimate the magnitude of the in-plane magnetization at $t = 2 \tau$.
A simple geometric argument is not possible for the inplane magnetization, but the exact treatment yields $\bar{M}^x = 0$ and $\bar{M}^y = (\alpha_{xy}/2) \sin^3 \theta$ (see SM).
Because each spin is acted upon by $I^z$ from $B_z$, and $I^x$ and $I^y$ from the interaction, behavior beyond a simple frequency shift is expected. 
The multi-peak behavior is most pronounced when the magnitude of the in-plane magnetization is largest, e.g. near $\theta = 90^\circ$.
There is also an S-function shift caused by the in-plane interaction, with magnitude $(\alpha_{xy}/4)$, which is due to the weak $\alpha_{xy}$ torque applied to the $z$-component of the spins after the $2\theta$ pulse (see SM).
For the third case where $\alpha_{xy} \neq \alpha_z$ (Fig.~\ref{fig:ratio_sweep}b) the S-function shifts from $\alpha_z$ and $\alpha_{xy}$ combine linearly, and the presence of $\alpha_z$ does not remove the multiple peaks generated by $\alpha_{xy}$.

Inverting the argument for the results of Fig.~\ref{fig:ratio_sweep}, in a laboratory setting the pulse variation experiment could be performed under different chosen directions for the $B_z$ fields relative to the sample's crystaline axis.
If the nuclei-nuclei coupling is mostly isotropic, the resulting NMR signals should not depend on the placement of the $z$-axis (in the absence of any other effects).
If the coupling is stronger along one axis than the other two, a clear S-function like that of Fig.~\ref{fig:ratio_sweep}a will occur along a specific direction of the applied field, while if it is weaker along one axis, an inverted S-function with severe hole-burning-like features should occur (Fig.~\ref{fig:ratio_sweep}c).

Fig.~\ref{fig:ratio_sweep}b includes experimental spectra~\cite{Mitrovic2008} from a superconducting phase, as a function of the applied $\theta$ pulse for qualitative comparison to our model (the frequency axis has been shifted and rescaled identically across all curves). At low power the peaks are at a low frequency, but as the power increases they shift to higher frequencies and show unusual non-monotonic behavior, matching the S-function fairly well.
Mapping power (dB) to a pulse angle ($^\circ$) is nearly impossible in experiment, especially when we predict large reductions in the signal near $90^\circ$, so more theoretical and experimental work is necessary.
Still, our preliminary model shows promise in explaining this unusual NMR measurement, originally ascribed to heating.
The evolution of the echo shape and position as a function of the pulse power and orientation of the applied field may permit us to reverse engineer details of this material's electronic spin susceptibility.

We now consider competition between the spin-susceptibility term and an NMR shift from pulse-induced heating.
Assuming that shift is monotonic and that the heating power follows the RF amplitude, $\theta^2$, its role can be distinguished from those of electronic correlations if $\theta$ is observed over most of a $2 \pi$ cycle.
In the case of a monotonic but non-linear shift (such as a shift associated with a phase transition) a similar technique can still be employed as long as a significant portion of the $2 \pi$ cycle is accessible experimentally.

\begin{figure}[htbp]
\centering
\includegraphics[width=\linewidth]{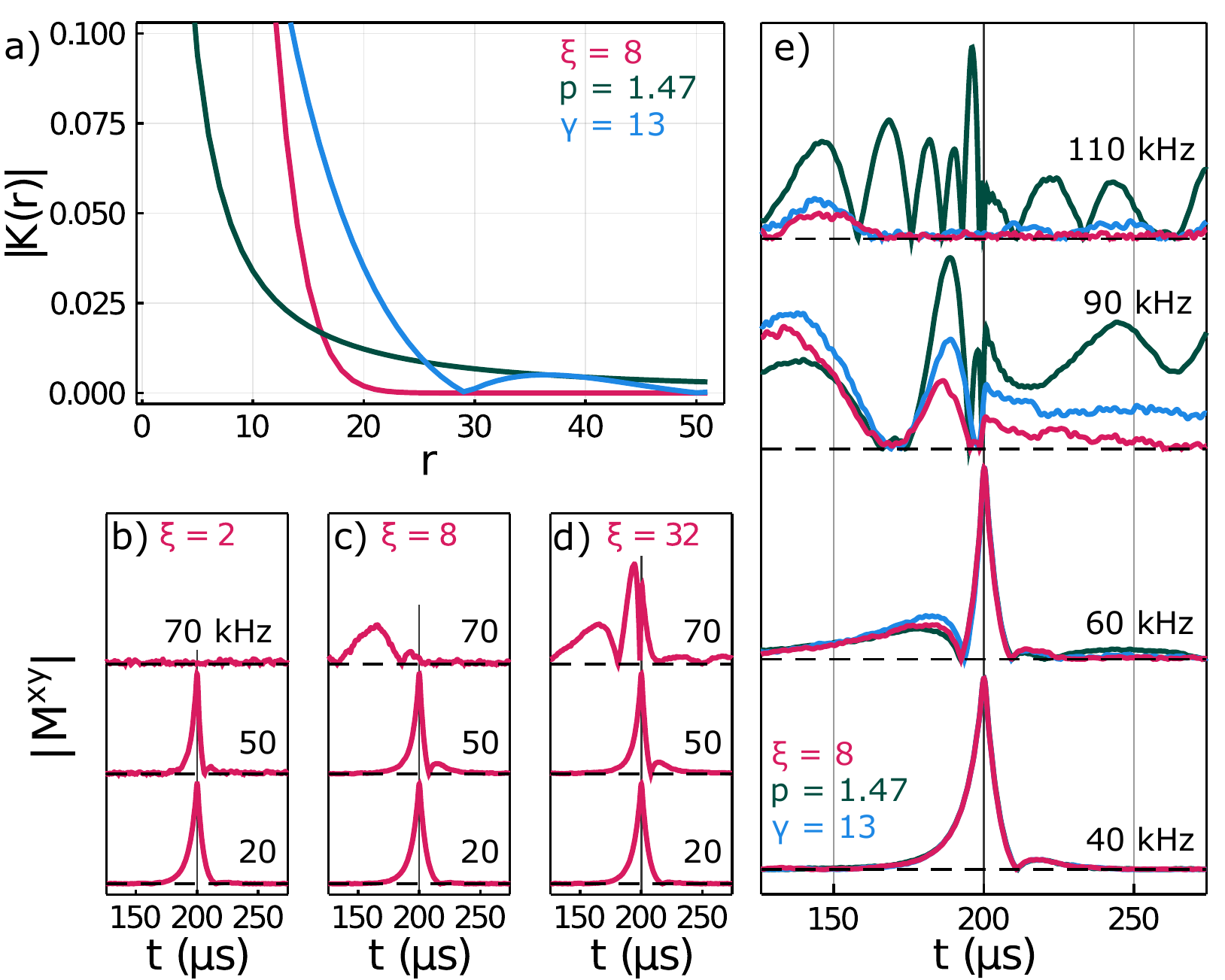}
\caption{
\textbf{a)} Absolute value of the three magnetization kernels, $K$: Gaussian (red) with $\xi = 8$ , power (green) with $p = 1.47$, and RKKY (blue) with $\gamma = 13$.
\textbf{b-d)} Time-domain spin echos for the Gaussian kernel with a short, medium, and long-range $\xi$.
\textbf{e)} Simulated time-domain spin echos for the kernels with medium length scale as given in \textbf{(a)} for increasing interaction strength. All three curves are identical for $\alpha_\textrm{eff} = 40$ kHz.
In \textbf{b-e)}, $\theta = 90^\circ$ and the total effective interaction, $\alpha_\textrm{eff}$ in \mbox{Eq.~\ref{aEff}}, is given in black on the right of each set of simulated spin echo curves.}
\label{fig:func_sweep}
\end{figure}

Furthermore, our work allows for the determination of the spatial extent of electron-electron correlations.
In real materials, each spinor will feel a local contribution from nearby nuclei, not a global average of the magnetization.
Between different materials and quantum spin phases, the type of radial decay in the susceptibility and its characteristic correlation length will vary.
To investigate this variation, we define the local magnetization $\vec{M}_i$ felt by a nucleus at site $r_i$ as the sum 
\begin{equation}
\vec{M}_i = \sum_j K(r_{ij}) \braket{\vec{I}_j}
\end{equation}
with $K$ a radial kernel for the interaction.
We study three choices of $K$ here.
First, a short-range Gaussian that depends on a correlation length $\xi$, $K(r) = e^{-\left(r/\xi \right)^2}$, motivated by the susceptibility expected from a gapped spin excitation.
Second, a long-range form given by a power $p$, $K(r) = r^{-p}$, motivated by a gapless spin excitation.
Finally, the RKKY form expected from spin interactions in a simple metal~\cite{Ruderman1954,Yosida1957} which is also dependent on a length $\gamma$, $K(x) = x^{-4} \left( x \cos{x} - \sin{x} \right)$ for $x = 2(r/\gamma)$.
In Fig.~\ref{fig:func_sweep}a the three functional forms for $K$ are plotted using parameters that yield similar length scales, for comparison.

Spin echo results identical to those of Fig.~\ref{fig:power_sweep} and Fig.~\ref{fig:ratio_sweep} are possible in this more realistic model if $\alpha$ and the length-scale parameter, $\xi$, $p$, or $\gamma$ are chosen appropriately (see SM).
The key parameter is the average effective interaction
\begin{equation}
\label{aEff}
\alpha_\textrm{eff} \equiv \sum_{d=x,y,z} \frac{\alpha_d}{3}  \sum_{ij} K(r_{ij})
\end{equation}
which is an integral of the interaction over the lattice and averaged over the three spin-spin spatial dimensions.
We find that a local interaction produces similar echoes to that of the global magnetization studied earlier, as long as the $\alpha$ used in the global case is similar to $\alpha_\textrm{eff}$ of the local one and the range of the interaction is sufficiently long.
In Fig.~\ref{fig:func_sweep}b-d, simulations with a gaussian functional for three values of $\xi$ are shown.
At small values of $\xi$, the spin echo acts similarly to the infinite-range model for weak $\alpha_\textrm{eff}$.
But as the interaction increases, the coupling to neighboring spins becomes so strong that extreme variations in the local effective magnetization occur throughout the lattice, destroying the echo.
For intermediate values of $\xi$ similar behavior is observed, but now the critical $\alpha_\textrm{eff}$ for complete destruction of an echo is larger.
When the correlation length is much larger than the lattice parameter ($\xi = 32$), the echo is identical to the results of the infinite-range coupling even for large $\alpha_\textrm{eff}$.
Therefore, the effective strength of the coupling ($\alpha_\textrm{eff}$) can be determined if an echo occurs, and the higher its value the longer range the electronic spin-spin correlations must be.
In a three-dimensional lattice, the critical minimal value of $\xi$ for a given $\alpha_\textrm{eff}$ will be smaller but will still extend beyond nearest-neighbor coupling.

Echos caused by interactions with similar $\alpha_\textrm{eff}$ but different radial forms are shown in Fig.~\ref{fig:func_sweep}e.
We see that although all three curves show similar qualitative trends, there are small details that distinguish them.
For example, in the pre-echo shoulder ($t = 150$ $\mu$s) the RKKY form always has the highest $\bar{M}^{xy}(t)$ value, followed by the gaussian, and then the power form.
Similarly, in the post-echo shoulder ($t = 250$ $\mu$s), the power form yields the largest $\bar{M}^{xy}(t)$ and the gaussian form the smallest.
At the largest $\alpha_\textrm{eff}$ value in Fig.~\ref{fig:func_sweep}e, these trends no longer hold because the echoes have disappeared for the gaussian and RKKY forms.
The RKKY form is an oscillating power-law decay, with its nodes partially cancelling long-range contributions and making it act like a short range interaction in our model.
The power law balances local versus average magnetization and prevents a complete breakdown of the spin echo phenomena.
Application of this method is particularly useful to deduce the spatial extent of electron-electron correlations through phase transitions.


Careful evaluation of NMR responces provides valuable insight into systems with complex emergent ordering.
Recent revisiting of NMR experiments on strongly correlated superconductors have shown that the assumed Knight shift in the superconducting phase can be highly dependent on the pulse strength applied in the NMR protocol~\cite{Ishida2019, Pustogow2019}.
We demonstrate that dependency of NMR shifts on the pulse power could be caused not only by external effects quenching an electronic phase, but also by the susceptibility of that electronic phase itself.
This encourages careful consideration of any NMR result in a strongly correlated system.
Clear signatures of nematic (anisotropic) ordering can be revealed by changing the pulse strength and orientation of the applied field.
Moreover, the methodology developed here  gives insight into the radial form and range of electronic correlations.
We hope that extensions of this work can ultimately lead to the ability to reverse engineer the full electronic susceptibility from simple NMR spectral measurements.

\begin{acknowledgments}
We thank Mladen Horvatić for helpful comments. This work was supported by the National Science Foundation under grant No. OIA-1921199. The calculations were conducted using computational
resources and services at the Center for Computation and Visualization, Brown University.
\end{acknowledgments}

\bibliography{refs}

\end{document}


\title{Supplementary Material for ``Signatures of electronic correlations and spin-susceptibility anisotropy in nuclear magnetic resonance''}

\author{Stephen Carr}
\affiliation{Department of Physics, Brown University, Providence, Rhode Island 02912-1843, USA}
\affiliation{Brown Theoretical Physics Center, Brown University, Providence, Rhode Island 02912-1843, USA.}
\author{Charles Snider}
\affiliation{Department of Physics, Brown University, Providence, Rhode Island 02912-1843, USA}
\author{D. E. Feldman}
\affiliation{Department of Physics, Brown University, Providence, Rhode Island 02912-1843, USA}
\affiliation{Brown Theoretical Physics Center, Brown University, Providence, Rhode Island 02912-1843, USA.}
\author{Chandrasekhar Ramanathan}
\affiliation{Department of Physics and Astronomy, Dartmouth College, Hanover, NH 03755, USA}
\author{J. B. Marston}
\affiliation{Department of Physics, Brown University, Providence, Rhode Island 02912-1843, USA}
\affiliation{Brown Theoretical Physics Center, Brown University, Providence, Rhode Island 02912-1843, USA.}
\author{V. F. Mitrovi\ifmmode \acute{c}\else \'{c}\fi{}}
\affiliation{Department of Physics, Brown University, Providence, Rhode Island 02912-1843, USA}
\date{\today}

\maketitle

\section{The nuclear spin ensemble}
The ensemble of nuclear spins is represented by a list of frequencies ($\nu_i$), weightings ($P_i$), positions ($r_i$), and initial states ($\psi_i^0$).
The frequencies $\nu_i$ are sampled from a Lorentz (Cauchy) distribution $C$ with linewidth $\Gamma$ and central frequency $\nu_0$, with weighted probablity according to
\begin{equation}
C(\nu) = \frac{\Gamma}{\pi} \frac{1}{(\nu - \nu_0)^2 + \Gamma^2}
\end{equation}
The weightings $P_i$ will be discussed in the next section.
We will assume $\psi_i^0 = \ket{\uparrow}$, i.e. all spins are initially in a perfect eigenstate with the maximum allowed $z$-component of spin.
For computational simplicity we take all nuclear spins to be spin-$1/2$, although we expect our findings to generalize to larger values as we are using a mean-field (classical) approach.

For future application of dissipation (requiring non-unitary operators) or beyond mean-field treatment, our simulations operate within the Linbladian paradigm for open systems \cite{Lindblad1976, Gorini1976, AmShallem2015}.
We use density matrices for each spinor, $\rho_i = \psi_i \psi_i^\dagger$, and follow a standard convention for converting the $2 \times 2$ Hamiltonians acting on the $2 \times 1$ states $\psi_i$ to $4 \times 4$ Linblad operators ($\hat{L}$) acting on the (vectorized) $4 \times 1$ density matrices $\rho_i$.
Time propagation of each spin in the ensemble is performed with a complex matrix exponential
\begin{equation}
    \rho_i(t+\Delta t) = e^{-i \hat{L}_i(t) \Delta t} \rho_i(t) e^{i \hat{L}_i(t) \Delta t}
\end{equation}
with $\hat{L}_i(t)$ the spin's Linbladian operator (generated from the Hamiltonians defined in the main text) and $\Delta t$ the time-step parameter of the simulation.
The magnetization of each spin in direction $\hat{n}$ is then evaluated as $\langle I^{\hat{n}}_i \rangle = \textrm{Tr}(\rho_i I^{\hat{n}} )$.
As we do not use any dissipation operators in this work, we have avoided discussion of the Linbladian method in the main text, and all results would be identical if we worked with Hamiltonians and state vectors $\psi_i$ instead. 
For the rest of the SM, we will simply define and work within the context of the state vectors $\psi_i$, as the conversion of all operators and states to Linbladians and density matrices adds no additional insight.

Since we use a numerical approximation of the matrix exponential, the appropriate value of the $\Delta t$ is determined by considering the maximum value obtained by the exponential's argument, which is proportional to $H_i(t) \Delta t $.
As such, if we choose (for simplicity) to select a global value for the time-step, it should be based on the convergence properties of the model at the largest considered $\alpha$ value.
We select $\Delta t = 0.15$ $\mu$s based on analysis of large $\alpha$ simulations.
The convergence of the model with respect to $\Delta t$ will be discussed in greater detail in an accompanying computationally-focused paper \cite{Snider2021}. The code is available in Ref.~\cite{SpinEchoCode}.

The spin echo pulse sequence is introduced with rotation operators, caused by a resonant RF pulse along a chosen direction.
We make the approximation that the pulse duration is smaller than $\tau$ and $\Delta t$, such that our pulses are instantaneous and given by the operator:
\begin{equation}
        R^d_\theta = e^{-i \theta I^d}.
\end{equation}

The time difference between the first pulse, $U_{90^\circ}$ ($\theta = \pi/2$), and second pulse, $U_{180^\circ}$ ($\theta = \pi$), is indicated by the variable $\tau$.

The simulations are performed in the rotating frame of the central frequency of the spin distribution, $\nu_0$.
In this frame, the evolution of each nuclear spin in the absence of interactions is given by 
\begin{equation}
    H_i^0 = -(\nu_i - \nu_0) I^z.
\end{equation}
This provides a more numerically stable treatment of both the ``fast'' time-scale of the spin precision and ``slow'' time-scale given by the pulse sequence $\tau$.
If the nuclear coupling is only strong between a specific magnetization direction in the $xy$-plane, such as a term like $M^x I_y$, then this strong interaction term will need to be rotated at the ``fast'' rate of $\nu_0$. This likely integrates out the effect into one more similar to the planar term $\alpha_{xy}$.

We have absorbed units of $\hbar$ and the gyromagnetic ratio $\gamma$ into the values of $\nu$ and $\alpha$, such that the only units are those of time ($t$,$\Delta t$,$\tau$) or frequency ($\nu$,$\Gamma$,$\alpha$).
For all our simulations, the central frequency $\nu_0$ is 10 MHz with a linewidth $\Gamma = 25$ kHz, $\tau$ is fixed at 100 $\mu$s, and $\Delta t = 0.15$ $\mu$s.

\section{The infinite-range approximation}

We introduced in the previous section the spin-dependent weightings $P_i$.
They provide an additional method of tuning the definition of the local magnetization $\vec{M}_i$, as follows:

\begin{equation}
    \vec{M}_i = \sum_{j \neq i} P_j K(r_{ij}) \braket{\vec{I}_j}.
\end{equation}

To most closely mimic a realistic system, one would assign each nucleus of the material its own frequency.
In studying the effect of the shape and range of the spin-spin interaction using $N$ total spins, this corresponds to taking equal weightings for each spin ($P_i = 1$) and each frequency $\nu_i$ sampled randomly from a Lorentz distribution.
This model allows for a direct analysis of how the spatial form and effective correlation length for the $\alpha$ interaction affects $M^{xy}(t)$ (Fig. 4 of the main text).
However, we find that large numbers of spin ($N > 10^5$) are necessary to converge the $M^{xy}(t)$ profiles at large values of $\alpha_\textrm{eff}$, as fine sampling of the Lorentizian distribution of $\nu$ becomes necessary. This means each calculation can take hours even when using a dedicated cluster-grade GPU.

This computational complexity encourages the use of an alternate strategy, where we instead consider an integrated sampling of the initial spin distribution, with the frequencies $\nu_i$ sampled evenly across the bandwidth and with each $P_i$ corresponding to the expected value of the distribution at that frequency.
We find excellent convergence with this approach. However, since each simulated spin corresponds to a statistical average instead of a physical spin in a real system, the implementation of spatial-dependence to the interaction becomes difficult, if not impossible. Therefore, we only use this weighted approach to the spin distribution for the infinite-range simulations (Fig. 1 and Fig. 2 of the main text).

For the infinite-range model, we perform our calculation on $40 \times 40 = 1,600$ spins, using the above $P_i$ weighting scheme to integrate the Lorentizian distribution.
For studying the effects of the functional form and range of the interaction, we use $400 \times 400 = 160,000$ spins with equal weightings.

\section{The Fourier Transform}

To convert our time-dependent complex-valued signals, $\bar{M}^{xy}(t) = \bar{M}^x(t) + i \bar{M}^y(t)$ into a spectrum comparable to those seen in NMR experiments, we perform the following processing steps.
First, we truncate $\bar{M}^{xy}(t)$ within a window centered at the spin echo and of size $2\tau$: the range $t = [\tau, 3 \tau] = [100,300]$ $\mu$s.
We then apply an envelope function to this window, to ensure $\bar{M}^{xy}(t)$ goes smoothly to zero at the boundaries:
\begin{equation}
\begin{split}
    f_e(t) &= \left[ \left( 1 + e^{-w (t_0 + (t - 2 \tau))} \right) \left( 1 + e^{-w (t_0 - (t - 2\tau))} \right) \right] ^{-1} \\
    \bar{M}^{xy}_e(t) &= f_e(t) \bar{M}^{xy}(t)
\end{split}
\end{equation}
and set $w = 100$ kHz and $t_0 = 20$ $\mu$s.
We then pad $32$ additional zeroes on both sides of $\bar{M}^{xy}_e(t)$ to help smooth out the resulting Fourier transform ($\bar{M}^{xy}_{e,z}(t)$).
Next, to peform the FFT we first apply an FFT-shift (to put the spin echo at the origin of the padded data), then a 1D FFT, and then another FFT-shift (to put $\omega = 0$ at the center of the returned frequency domain).
Finally, we add 10 MHz to the resulting frequency axis to account for the rotating frame and plot the absolute value of the spectrum, $|\tilde{M}^{xy}_{e,z}(\omega)|$.

\section{Interaction-free Magnetization}

Here we derive the average magnetization of the spin ensemble as a function of time $t$ and pulse strength $\theta$ in the case of a non-interacting system ($\alpha = 0$).
This is useful in understanding the effects of the spin-spin interaction, as it gives a good estimation of $M_i$ during the spin echo.
Working in the rotating frame, we can assume the spins are distributed symmetrically about $\nu = 0$.
We will consider first a single spin, with resonant frequency $\nu$, and derive $\braket{I^d}$ for $d = x,y,z$.
Then, we will estimate the mean magnetization of the entire ensemble by considering an average over a typical distrubtion of $\nu$.

We begin the single-spin case by defining the spin operators (with $\hbar = 1$) as

\begin{equation}
    I^x = \frac{1}{2} \begin{pmatrix} 0 & 1 \\ 1 & 0 \end{pmatrix},
    I^y = \frac{1}{2} \begin{pmatrix} 0 & -i \\ i & 0 \end{pmatrix},
    I^z = \frac{1}{2} \begin{pmatrix} 1 & 0 \\ 0 & -1 \end{pmatrix}
\end{equation}

Treating the spin-$1/2$ wavefunction as

\begin{equation}
    \psi = \begin{pmatrix} \psi^A \\ \psi^B \end{pmatrix}
\end{equation}

we have

\begin{equation}
\begin{split}
    M^x &= \bra{\psi} I^x \ket{\psi} = \textrm{Re}(\bar{\psi^A} \psi^B) \\
    M^y &= \bra{\psi} I^y \ket{\psi} = \textrm{Im}(\bar{\psi^A} \psi^B) \\
    M^z &= \bra{\psi} I^z \ket{\psi} = \frac{1}{2} \left( |\psi^A|^2 - |\psi^B|^2 \right)
\end{split}
\end{equation}

The $\theta$-strength $x$-pulse and the $2 \theta$-strength $y$ pulse are given by the operators

\begin{equation}
    \begin{split}
        R_\theta^x &= e^{-i \theta I^x} =
        \begin{pmatrix}
        \cos \frac{\theta}{2} & -i \sin \frac{\theta}{2}  \\
        -i \sin \frac{\theta}{2}  & \cos \frac{\theta}{2} 
        \end{pmatrix} \\
        R_{2\theta}^y &= e^{-i 2\theta I^y} =
        \begin{pmatrix}
        \cos \theta & \sin \theta \\
        -\sin \theta & \cos \theta
        \end{pmatrix} \\
    \end{split}
\end{equation}

and time propagation by $\tau$ is given by the operator

\begin{equation}
    O_\tau = e^{-i (-\nu I^z) \tau} =
    \begin{pmatrix}
    e^{i \phi} & 0 \\
    0 & e^{-i \phi} 
    \end{pmatrix}
\end{equation}

with $\phi = \tau \nu / 2$

Our spinor of frequency $\nu$ begins in the up-spin state at $t_0 = 0$:

\begin{equation}
    \psi(0) = \begin{pmatrix} 1 \\ 0 \end{pmatrix}
\end{equation}

and its state at the time of spin echo is given by

\begin{equation}
    \begin{split}
    \psi(2\tau) &= O_\tau R_{2 \theta}^y O_\tau R_\theta^x \psi(0) \\
     &= \begin{pmatrix}
     \cos \theta \cos \frac{\theta}{2} e^{2 i \phi} - i \sin \theta \sin \frac{\theta}{2}\\
     - \sin \theta \cos \frac{\theta}{2} - i \cos \theta \sin \frac{\theta}{2} e^{-2 i \phi} 
     \end{pmatrix}
    \end{split}
\end{equation}

such that

\begin{equation}
    \begin{split}
        \bar{\psi^A} \psi^B &= \frac{i}{2} \sin^3 \theta - \cos^2 \theta \sin \theta e^{-2i \phi} \left(1 + \frac{i}{2} e^{-2i \phi} \right) \\
        |\psi^A|^2 &= \sin^2 \frac{\theta}{2} + \cos^3 \theta - \cos \theta \sin^2 \theta \sin 2\phi \\
        |\psi^B|^2 &= \cos^2 \frac{\theta}{2} - \cos^3 \theta + \cos \theta \sin^2 \theta \sin 2\phi
    \end{split}
\end{equation}

and yielding expectation values of

\begin{equation}
    \begin{split}
        M^x(2 \tau) &= - \cos^2 \theta \sin \theta \left( \cos 2\phi + \frac{1}{2} \sin 4\phi \right) \\
        M^y(2 \tau) &= \frac{1}{2} \sin^3 \theta - \cos^2 \theta \sin \theta \left( \frac{1}{2} \cos 4\phi + \sin 2\phi \right) \\
        M^z(2 \tau) &= \frac{1}{4} \left( \cos \theta + \cos 3\theta \right) - \cos \theta \sin^2 \theta \sin 2\phi
    \end{split}
\end{equation}

To estimate the effective average magnetization of the entire ensemble, $\vec{\bar{M}}$, we first note that in almost all nuclear systems the spin distribution is symmetrical about the central frequency, and so any term odd in $\phi$ (e.g. sin) will go to zero.
For the even terms in $\phi$ (e.g. cos), we integrate $\cos (m \phi) = \cos (m \tau \nu / 2)$ against the spins' Lorentz distribution of linewidth $\Gamma$:

\begin{equation}
    \int_{-\infty}^{\infty} d\nu \frac{\Gamma}{\pi} \frac{\cos(\beta \nu) }{\nu^2 + (\Gamma)^2} = e^{- \Gamma \beta} = e^{-m \Gamma \tau / 2}
\end{equation}

giving

\begin{equation}
    \begin{split}
    \bar{M^x}(2\tau) &= - e^{- \Gamma \tau} \cos^2 \theta \sin \theta \\
    \bar{M^y}(2\tau) &= \frac{1}{2} \left( \sin^3 \theta - e^{-2 \Gamma \tau} \cos^2 \theta \sin \theta \right)\\
    \bar{M^z}(2\tau) &= \frac{1}{4} \left( \cos \theta + \cos 3\theta \right).
    \end{split}
\end{equation}

When $\bar{M^x}$ is largest $(\theta = 45^\circ$) any enhancement in $\bar{M^x}$ at small $\tau$ is compensated by a decrease in $\bar{M^y}$.
There is some particular $\tau$ where $\bar{M^x} = \bar{M^y}$ during the spin-echo.
But such tuning is unlikely to be useful to understanding the material properties as we are working in the rotating frame.
The $M^x$ and $M^y$ corresponding to particular crystal axes are rotating at $10$ MHz relative to those here, and thus it is likely that these effects are averaged out and effectively remove any obvious $\tau$ dependence.
There may be some interplay between the $x$ and $y$ component of the spins due to the interaction, and thus some effect of $\tau$ is expected, but it is beyond the scope of our analysis here.

In our simulations, $\Gamma = 25$ kHz and $\tau = 100$ $\mu$s, so $\cos(2\phi) \to 0.0821$ and $\cos(4\phi) \to 0.0067$.
We can approximate $\cos(2 \phi) \approx \cos(4 \phi) \approx 0$ and arrive at $\bar{M^x}(2 \tau) \approx 0$ and $\bar{M^y}(2 \tau) \approx (1/2) \sin^3 \theta$, in good agreement with the spectral response to $\alpha_z$ and $\alpha_{xy}$ in Fig. 2 of the main text.

\section{Comparison of infinite-range to finite-range interaction}

Here we compare the NMR signals of the infinite-range anisotropic model and those of an equivalent finite-range Gaussian interaction with $\xi = 32$.
We find that for $\alpha_\textrm{eff} \leq 120$ kHz, good agreement is found between the two forms of the interaction.

\begin{figure}
\centering
\includegraphics[width=\linewidth]{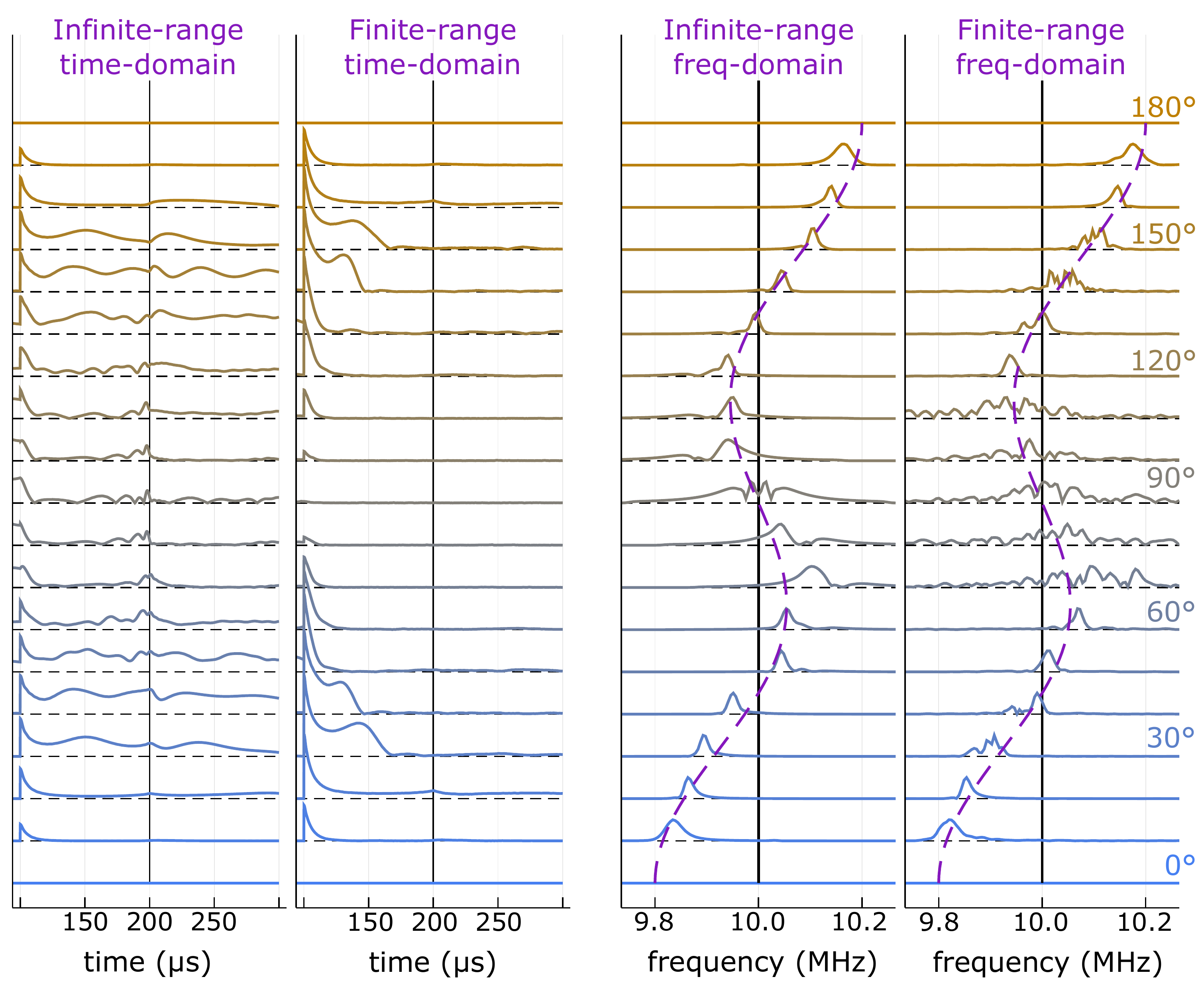}
\caption{Comparison of simulation results between the anisotropic infinite-range interaction in Fig. 2b of main text and a finite-range Gaussian interaction with $\xi = 32$ and equivalent effective $\alpha_z$ and $\alpha_{xy}$ strength.
The average in-plane magnetization is shown in both the time and frequency domain for different pulse angles $\theta$, with the anisotropic coupling $\alpha_z = 500$ kHz and $\alpha_{xy} = 100$ kHz. Note that the finite-range time-domain signal (second panel) has been multiplied by a factor of $2$, but still no signal at $t = 200$ $\mu$s is visible in this case.}
\label{fig:aniso_compare_b}
\end{figure}

First we consider the $[\alpha_z$, $\alpha_{xy}] = [500,100]$ kHz from Fig. 2b of the main text.
In Fig. \ref{fig:aniso_compare_b} we show both the time and frequency NMR signals for both types of interaction.
At $\theta$ near $0^\circ$ and $180^\circ$, the agreement is already poor, but as $\theta$ approaches $90^\circ$ the finite-range model shows the echo breakdown behavior noted in the main text's Fig. 4.
$\alpha_\textrm{eff}$ in this case is too large because of the large $\alpha_z$ value.
The S-function behavior in the NMR shift is still recovered, but the delicate structure of the frequency signal is lost when $\theta$ is close to $90^\circ$.

\begin{figure}
\centering
\includegraphics[width=\linewidth]{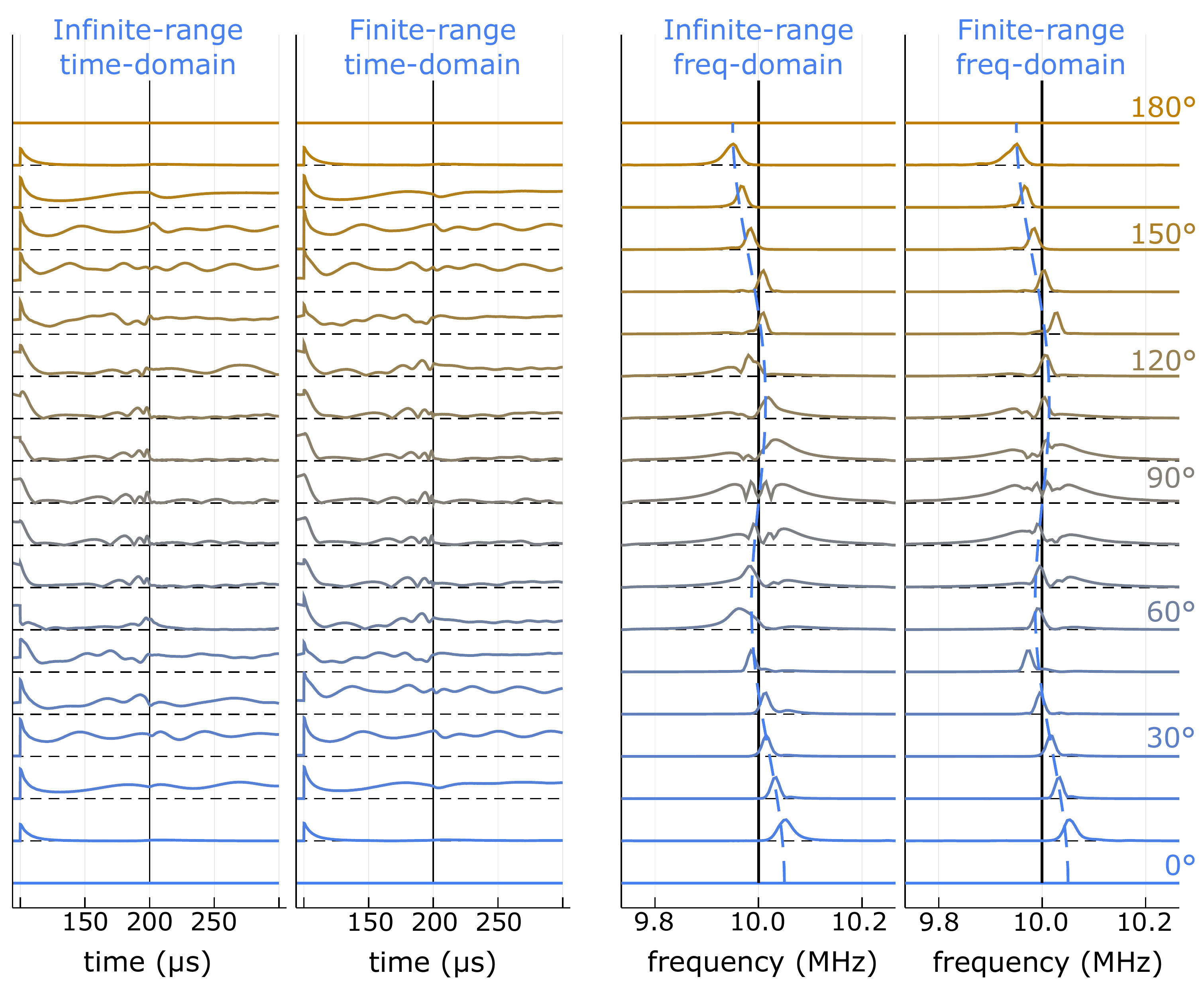}
\caption{Comparison of simulation results between the anisotropic infinite-range interaction in Fig. 2c of main text and a finite-range Gaussian interaction with $\xi = 32$ and equivalent effective $\alpha_z$ and $\alpha_{xy}$ strength.
The average in-plane magnetization is shown in both the time and frequency domain for different pulse angles $\theta$, with the anisotropic coupling $\alpha_z = 0$ kHz and $\alpha_{xy} = 100$ kHz.}
\label{fig:aniso_compare_c}
\end{figure}

If instead we take $[\alpha_z$, $\alpha_{xy}] = [0,100]$ kHz (Fig. 2c of the main text), the agreement near $\theta = 90^\circ$ is much improved, as shown in Fig. \ref{fig:aniso_compare_c}.
The time and frequency domain signals are in good agreement between the two types of interactions, showing that as long as the spin echo is not completely disrupted by electron mediated interactions, measurement of electronic nematicity will be possible.

\section{FID effects from interaction}

\begin{figure}
\centering
\includegraphics[width=\linewidth]{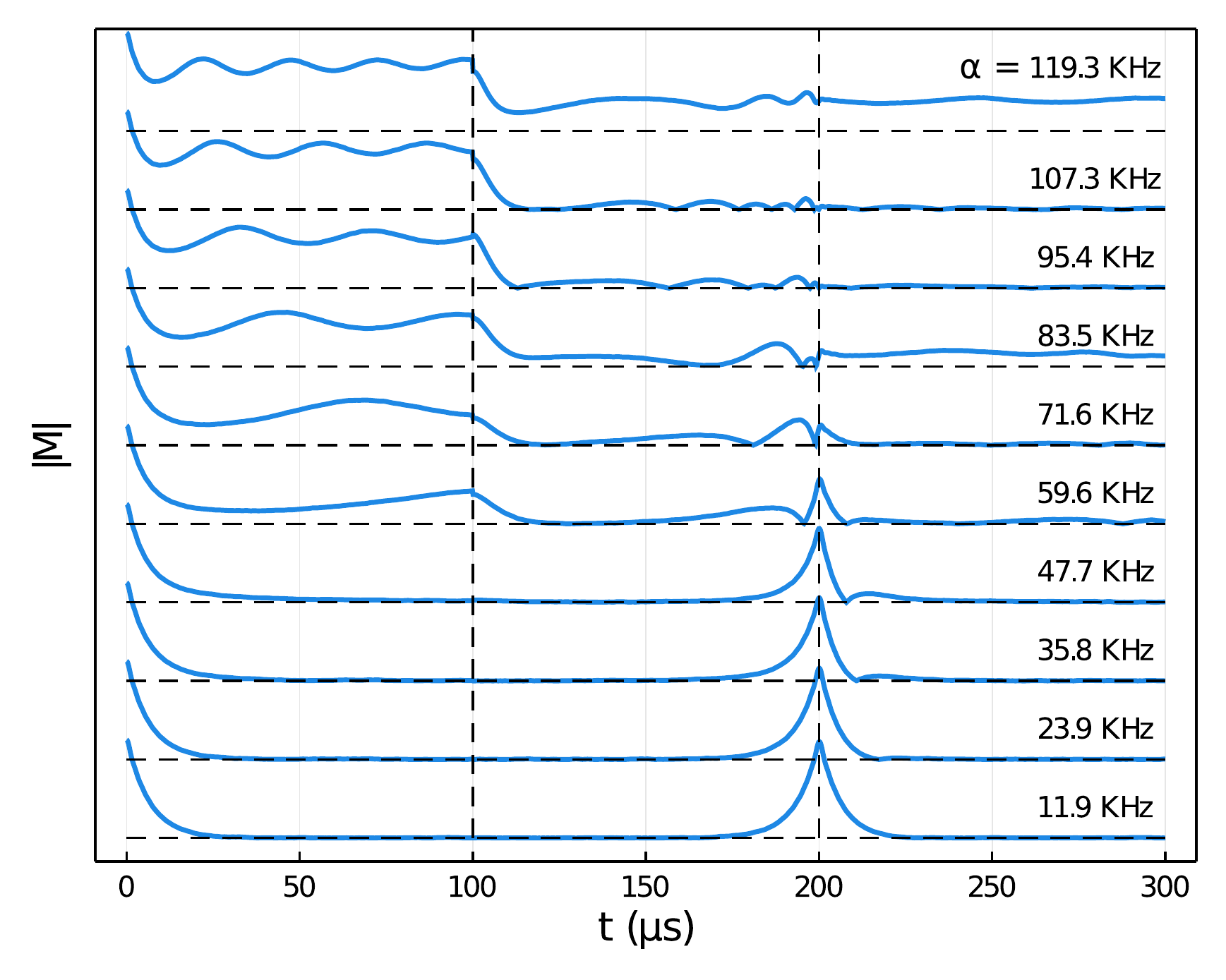}
\caption{In-plane magnetization for a Gaussian interaction with $\xi = 32$ and $\theta = 90^\circ$ for various values of $\alpha_\textrm{eff}$ (noted in black on right of each curve). A $180^\circ$ pulse is applied at $\tau = 100$ $\mu$s and the spin echo occurs at $2\tau = 200$ $\mu$s.}
\label{fig:FID_plot}
\end{figure}

In Fig. \ref{fig:FID_plot} time-domain spin echos are shown over the entire simulated time range, $t = [0,3 \tau]$ using a long-range ($\xi = 32$) isotropic Gaussian interaction. 
For $\alpha < 50$ kHz, the free induction decay (FID) shows the standard decay from the variance in spin resonances.
However at larger $\alpha$, the FID retains a finite, mildly oscillating, value.
The background of the FID, and its oscillation frequency, are both proportional to $\alpha$.
For $\alpha > 50$ kHz, the lack of decay in the FID also causes ringing in the $|M^{xy}|$ signal just after $\tau$, showing that the interaction not only modifies the spin echo shape, but also introduces new frequency components to the spectra of the FID as well.
We note that a larger or shorter $\tau$ does not remove these features from the FID, although the inclusion of $T_1$ or $T_2$ decay processes would lessen it with increasing $\tau$.
In the next section, we take a closer look on the structure and origin of the FID ringing.

\section{Shifts and ringing caused by \texorpdfstring{$\alpha_{xy}$}{axy}}

\begin{figure}
\centering
\includegraphics[width=\linewidth]{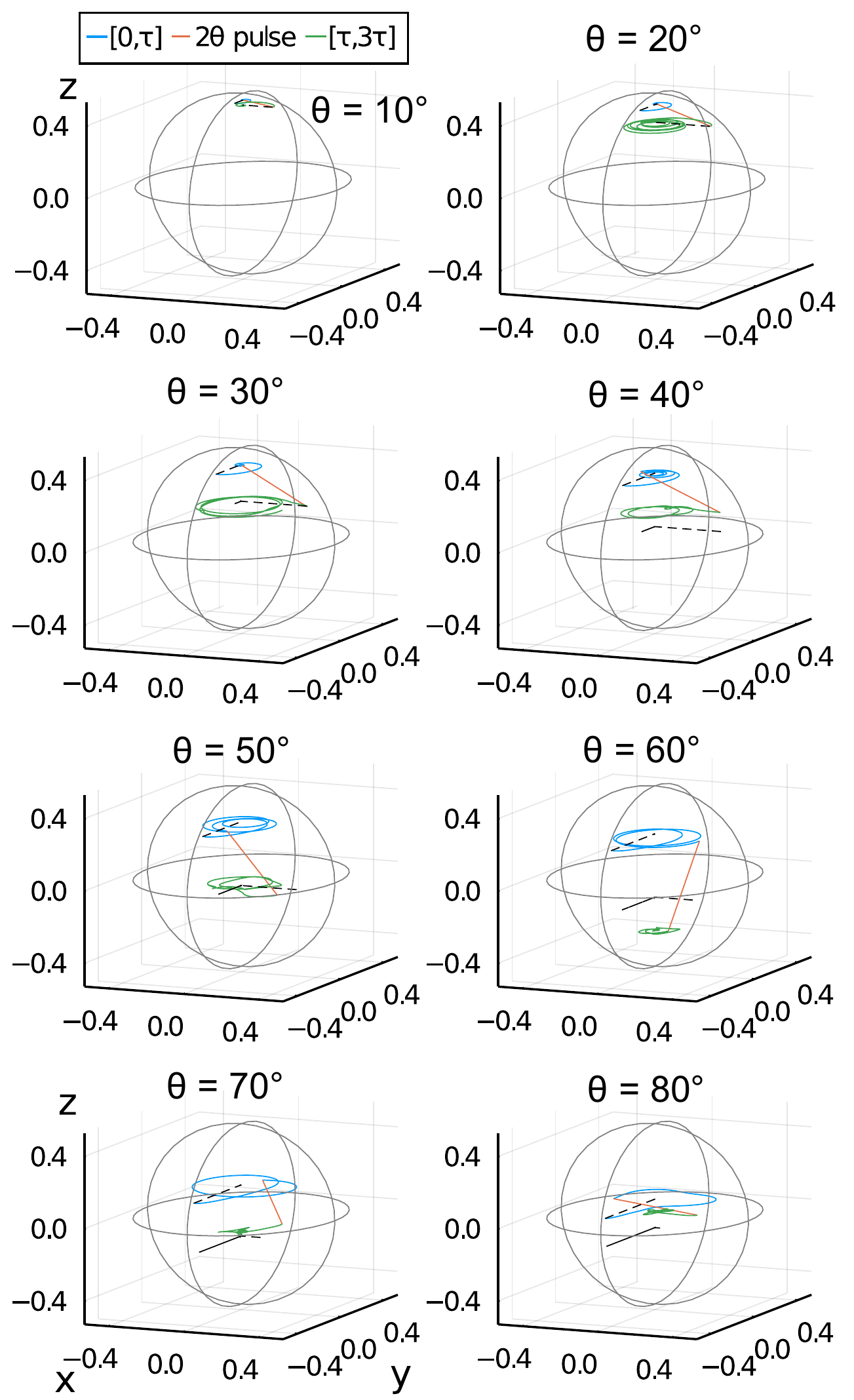}
\caption{
Three-dimensional plots of $\vec{\bar{M}} = [\bar{M}^x,\bar{M}^y,\bar{M}^z]$ for $\alpha_{xy} = 100$ kHz and $\alpha_z = 0$ in the infinite range approximation.
Each panel shows the result from a different applied pulse, $\theta$.
The motion of the average magnetization between the $\theta$ and $2\theta$ pulse is given in blue ($[0,\tau]$), the movement caused by the $2\theta$ pulse is given in orange, and the motion after the $2\theta$ pulse is given in green ($[\tau,3\tau]$).
The behavior in the absence of $\alpha_{xy}$ is given by the dashed black line in each case, with the $2\theta$ pulse line omitted.
Three grey circles outline $|\vec{\bar{M}}| = 1/2$ along each axis.
}
\label{fig:axy_plot}
\end{figure}

The complicated behavior of the FID for large $\alpha_\textrm{eff}$ and the S-function frequency shift caused by $\alpha_{xy}$ are both related to the interplay between the in-plane interaction and the pulse angle.
In Fig. \ref{fig:axy_plot}, the behavior of the average magnetization $\vec{\bar{M}}$ is shown as a function of pulse strength $\theta$, with and without an $\alpha_{xy} = 100$ kHz interaction.

Without an interaction (dashed black lines in Fig. \ref{fig:axy_plot}), $\vec{\bar{M}}$ begins at a negative $\bar{M}^y$ value and positive $\bar{M}^z$ value just after the $\theta$ pulse, and during the FID the $\bar{M}^y$ value decays to zero.
After the $2\theta$ pulse, the magnetization has a finite $M^x$ value, which then decays before an $M^y$ echo occurs at time $2 \tau$.
In contrast, the in-plane interaction (colored lines) causes $\vec{\bar{M}}$ to precess about the $z$-axis, and so the straight lines to $\bar{M}^{xy} = 0$ are replaced by spirals and circular orbits.
This is unexpected at first glance: these simulations have $\alpha_z = 0$, so why does the spin appear to precess about $\hat{z}$?

Consider the average magnetization's evolution in the $\theta = 30^\circ$ case just after the $\theta$ pulse (blue line).
The total magnetization is given by $\vec{\bar{M}} = (1/2)[0,-\sin \theta, \cos \theta]$, introducing an additional term of $(\alpha_{xy}/2) I^y \sin \theta$ in the Hamiltonian.
As the pulse is well below $90^\circ$ each spin has a large $z$ component, causing the $I^y$ operator to rotate the spins in the $\hat{x}$ direction.
After a small time $\Delta t$, the magnetization is $\vec{\bar{M}} \sim (1/2)[\sin \Delta t, -\sin \theta \cos \Delta t, \cos \theta]$, i.e. identical to its original value but slightly rotated around the $z$-axis.
This continues as the variance in the resonance frequencies $\nu$ cause attenuation of the signal, making the magnetization spiral into the $\bar{M}^{xy} = 0$ condition.
Similar behavior occurs after the $2 \theta$ pulse from the non-zero $\bar{M}^x$ value that occurs when $\theta \neq 90^\circ$, but in this case the in-plane signal does not decay as rapidly.

Because the spectrum during the spin echo (at $t = 2 \tau$) depends on this effective additional in-plane torque, we can estimate its precession frequency by taking the product of the $z$-axis magnetization and the in-plane magnetization: $\approx \alpha_{xy} (1/2)(\cos \theta + \cos 3 \theta) \bar{M}_{xy}$.
In the absence of the interaction, $|\bar{M}^{xy}(2 \tau)| \approx (1/2)\sin^3 \theta$, but because the interaction causes ringing in the FID, we find that this is no longer a good estimate even at small $\theta$ (see the $\theta = 20^\circ$ green curve: it never decays to zero after the $2\theta$ pulse).
Therefore, we make the rough approximation $|\bar{M}^{xy}(2 \tau)| \approx 1/2$, giving an effective spectral shift of $(\alpha_{xy}/4)(\cos \theta + \cos 3\theta)$, as used in Fig. 2 of the main text.

\clearpage

\bibliography{refs}